# Interpretation of EMF data in Ag-Au-S system.

Ya.I. Korepanov[1]

[1] D.S. Korzhinsky Institute of Experimental Mineralogy of the Russian Academy of Sciences (IEM RAS)

*e-mail: yakoff@iem.ac.ru*

Abstract

The EMF dependencies for Ag-Au-S systems were analyzed and it was found that the reactions described in reference [1] are inaccurate. Additionally, the assumption that the equilibrium state of Au remains unchanged during the measurement process was shown to be incorrect.

Based on the EMF data provided in White's 1957 paper, the compositions of the alloys involved in the reactions were determined. The reaction equations were then modified to include temperature dependencies of the alloy composition. The resulting equations are as follows:

$1/(1-x)Ag(cr) + Ag_3AuS_2(cr) = 2Ag_2S(cr) + 1/(1-x)Ag_xAu_{1-x}$, $x = +0{,}4535 - 3{,}513 * 10^{-4}\,T$

$1/(1-x)Ag(cr) + 2AgAuS(cr) = Ag_3AuS_2(cr) + 1/(1-x)Ag_xAu_{1-x}$, $x = 0{,}3092 - 3{,}341 * 10^{-4}\,T$

$1/(1-x)Ag(cr) + Au_2S(cr) = AgAuS(cr) + 1/(1-x)Ag_xAu_{1-x}$, $x = -0{,}0131 + 5{,}769 * 10^{-5}\,T$

The validity of these reaction equations was demonstrated, and the accurate thermodynamic functions for the phases were determined based on these equations.

*Keywords: Ag-Au-S system, silver, gold, sulfur, solid solution, alloy, geothermometer, EMF, thermodynamic properties, chemical potential.*

Gold-silver sulfides are not economically valuable compared to native gold, but they are common in gold-silver deposits. Studying their thermodynamic properties is important for understanding gold's geochemistry, transport, and deposition [1]. The Ag-Au-S system contains three stable sulfides at temperatures below 386 K: acanthite ($Ag_2S$), uytenbogaardtite ($Ag_3AuS_2$), and petrovskaite (AgAuS). However, Barton [2] noted uncertainties in the thermodynamic data, and Barton and Skinner (1979) reported an inconsistency in the standard Gibbs free energy of formation for metastable Au2S, which decomposes above 490 K.

The purpose of this article is to demonstrate how inaccuracies in recording reactions can lead to distorted data interpretation. The EMF method [3-4] is the most accurate and stable method for determining the thermodynamic parameters of the system in question. The article aims to obtain accurate thermodynamic data as described by Osadchy and Rappo [1] and show the impact of reaction equation accuracy on data consistency.

**Evaluation of EMF values from paper [1].**

The article [1] presents the following reactions:

$Ag(cr) + Ag_3AuS_2(cr) = 2Ag_2S(cr) + Au(cr)$

$Ag(cr) + 2AgAuS(cr) = Ag_3AuS_2(cr) + Au(cr)$

$Ag(cr) + Au_2S(cr) = AgAuS(cr) + Au(cr)$

These equations were written under the assumption that gold is stable in the Ag-Au-S system at experimental temperatures ranging from 310 K to 386 K in electrochemical experiments. Based on phase relations, determination of chemical potential [3], establishment of equilibrium at the measured boundary [4], and alloy data from White's article [5], the reaction equations were modified as follows:

$1/(1-x)Ag(cr) + Ag_3AuS_2(cr) = 2Ag_2S(cr) + 1/(1-x)Ag_xAu_{1-x}$, x=+ 0,4535-3,513*10$^{-4}$ T    R1

$1/(1-x)Ag(cr) + 2AgAuS(cr) = Ag_3AuS_2(cr) + 1/(1-x)Ag_xAu_{1-x}$, x=0,3093-3,341*10$^{-4}$ T    R2

$1/(1-x)Ag(cr) + Au_2S(cr) = AgAuS(cr) + 1/(1-x)Ag_xAu_{1-x}$, x= -0,0131+5,769*10$^{-5}$ T    R3

Since the measured value in the electrochemical cell corresponds to the difference in the chemical potentials of silver in the sample system and the comparison system, the reaction energy can be written as:

$$\Delta_r G^0 = -mzFE$$

Where m represents the number of ions involved in the reaction $m = \frac{1}{1-x}$ for all equations [3] and z represents the valence of the conducting ion in the electrolyte.

Using the EMF values provided in the article [1], the compositions of the alloy in ternary equilibria Tab.1. were calculated based on the equation:

$$-\frac{\Delta G_{Ag}(x)}{nF} - E = 0$$

**Table 1.** Values of EMF and compositions of the $Ag_xAu_{1-x}$ alloy from temperature for the phase associations studied in the article [1].

| $Ag_2S$-$Ag_3AuS_2$-$Ag_xAu_{1-x}$ | | | $Ag_3AuS_2$-$AgAuS$-$Ag_xAu_{1-x}$ | | | $AgAuS$-$Au_2S$-$Ag_xAu_{1-x}$ | | |
|---|---|---|---|---|---|---|---|---|
| T / K | E / mV | x | T / K | E / mV | x | T / K | E / mV | x |
| 312.6 | 106.25 | 0.34508 | 310.8 | 150.37 | 0.20591 | 307.6 | 298 | 0.00472 |
| 318.6 | 107.51 | 0.34068 | 313.2 | 150.78 | 0.20543 | 312.6 | 299 | 0.0049 |
| 322.4 | 108.11 | 0.34087 | 315.2 | 151.43 | 0.20417 | 317.9 | 299 | 0.00527 |
| 324.2 | 108.37 | 0.33908 | 318.2 | 152.2 | 0.20287 | 323.8 | 300 | 0.00552 |
| 330.6 | 109.47 | 0.33528 | 318.2 | 151.9 | 0.2037 | 330.2 | 300 | 0.006 |
| 333.2 | 110.14 | 0.33638 | 320.2 | 152.66 | 0.20215 | 335.6 | 301 | 0.00621 |
| 337.5 | 110.59 | 0.33576 | 322.6 | 153.08 | 0.20166 | 341.4 | 301 | 0.00667 |
| 344.4 | 112.11 | 0.33227 | 324 | 153.57 | 0.2007 | | | |
| 349.9 | 112.92 | 0.33081 | 324.6 | 153.61 | 0.20076 | | | |
| 355.2 | 114.05 | 0.32825 | 327.8 | 154.35 | 0.19962 | | | |
| 360.1 | 114.75 | 0.32592 | 331.2 | 155.03 | 0.19871 | | | |
| 366 | 115.93 | 0.3284 | 333.6 | 155.56 | 0.19793 | | | |
| 372.6 | 117.05 | 0.32082 | 335.8 | 156.14 | 0.19698 | | | |
| 376.6 | 117.81 | 0.32078 | 339.4 | 157.01 | 0.19466 | | | |
| 381 | 118.72 | 0.31884 | 342.6 | 157.76 | 0.19451 | | | |
| 385.8 | 119.54 | 0.31884 | 347.4 | 158.94 | 0.19141 | | | |
| | | | 350.6 | 159.5 | 0.1921 | | | |
| | | | 356.8 | 160.9 | 0.19014 | | | |
| | | | 361.4 | 162.12 | 0.18823 | | | |
| | | | 367.4 | 163.42 | 0.18653 | | | |
| | | | 371.8 | 164.55 | 0.18485 | | | |
| | | | 378 | 165.54 | 0.18401 | | | |
| | | | 382.4 | 166.69 | 0.18232 | | | |

To demonstrate that the reaction presented closely represents the actual processes occurring in nature and in the electrochemical cell, the phase diagram was simulated within the measured temperature range using TernApi programs [6]. The simulation yielded highly consistent results (Fig.1) in terms of the alloy composition, as determined through thermodynamic data (modeling), and as calculated from the chemical potential [3] and EMF dependencies of the cells obtained by the authors [1], with the EMF dependencies of the alloy from the White paper [2], using precise recording of the reaction.

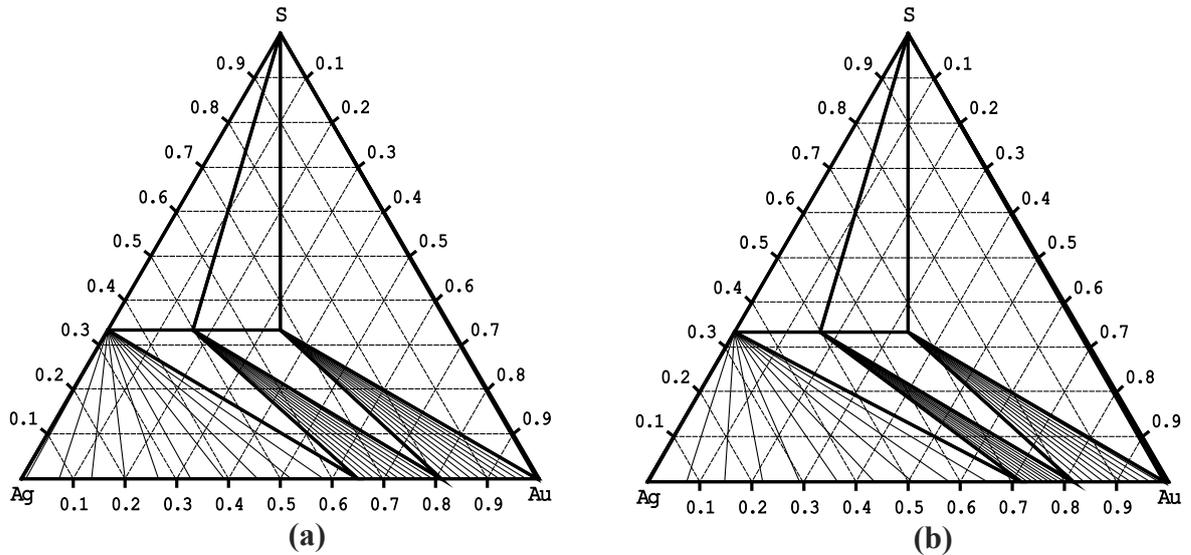

**Fig.1.** Modeling of the Ag-Au-S phase diagram. a - using new reaction equations, b – using reactions from the article [1]. For T=312.6 K in the phase association, the alloy $Ag_xAu_{1-x}$ x= 0.34508 was calculated. This value is in good agreement with the simulation results of 0.348 (less than 1%) (a) and goes against the simulation results of 0.284 (about 18%) (b).

Tables 2-4 provide a comparison between the published data and the values determined in this article.

**Table.2.** Standard thermodynamic functions at 298.15 K and 1 bar for reactions 1-3, calculated from experimental data given in Table 1

|  | $\Delta_r G^0$ (J/mol) | $\Delta_r H^0$ (J/mol) | $\Delta_r S^0$ (J/mol) |
|---|---|---|---|
| R1 | -15380± 270 | - -10180.27261± 270 | -17.44303 |
| [1] | –10010 ± 270 | –4830 ± 270 | 17.370 ± 0.700 |
| R2 | -18022± 280 | -12191± 280 | -19.558± 0.750 |
| [1] | –14240 ± 280 | –7650 ± 280 | 28022.100 ± 0.750 |
| R3 | -28817± 800 | - 25788± 800 | -10.16± 2.000 |
| [1] | –28700 ± 800 | –26190 ± 800 | 8.420 ± 2.000 |

**Table.3.** Coefficients a and b in equation $\Delta_f G^0(T) = a + bT$

| Phase | a (J/mol) | b (J/K·mol) |
|---|---|---|
| (Ag3AuS2, uytenbogaardtite) | -62772.42144 | -31.69013 |
| (AgAuS, petrovskaite) | -27480.07936 | -5.29712 |
| (Au2S, cr) | -1246.35069 | 2.93879 |

**Table.4.** Standard thermodynamic properties for crystalline phases in the Ag-Au-S system at 298.15 K and 1 bar

| Phase | $\Delta_f G^0$ (J/mol) | $\Delta_f H^0$ (J/mol) | $S^0$ (J/K*mol)S | Source |
|---|---|---|---|---|
| Ag(cr) | 0 | 0 | 42.55 ± 0.21 | 1 |
| Au(cr) | 0 | 0 | 47.49 ± 0.21 | 1 |
| S(rhomb) | 0 | 0 | 32.05 ± 0.05 | 1 |
| Ag2S, acanthite | –39700 ± 1000 | –32000 ± 1000 | 142.9 ± 0.3 | 1 |
| Ag3AuS2, cr, low (uytenbogaardtite) | –63440 ± 6300 | | | 2 |
| | –69478 ± 1200 | –57280 ± 1200 | 280.28 ± 0.2 | 4 |
| | -72220± 1200 | - 62247± 1200 | 207.5± 0.2 | 5 |
| AgAuS, cr, low (petrovskaite) | –22420 ± 4200 | | | 2 |
| | –27620 ± 1200 | –24800 ± 1200 | 131.56 ± 0.2 | 4 |
| | -29060± 1200 | -27090± 1200 | 116.79± 0.2 | 5 |
| Au2S(cr) | 10820 ± 8400 | | | 2 |
| | 28660 ± 10500 | | | 3 |
| | 1077 | 1390 | 128.1 | 4 |
| | -370 | -1246 | 129.96 | 5 |
| 1 - Robie and Hemingway [7]; 2 and 3 - calculated from Barton [8] and Barton and Skinner [9] respectively using data for S2(g) from Robie and Hemingway [7] ;4- Osadhii, Rappo [1] 5 - present study. (data 1-4 is a direct citation from [1]) | | | | |

## 4. Discussion.

From the presented graphs and tables, it is evident that accuracy in recording the reaction equations leads to significant discrepancies in the determined values. In Fig.2, the difference between the dependencies of the alloy composition (which essentially reflects the chemical potential of the element being measured in the system), situated within the triple association $Ag_2S$-$Ag_3AuS_2$-$Ag_xAu_{1-x}$, in the temperature determined through modeling using the TernApi program [6] is illustrated and compared with the values obtained by direct determination of the composition through the chemical potential, which accurately reflects the equilibrium composition of the alloy.

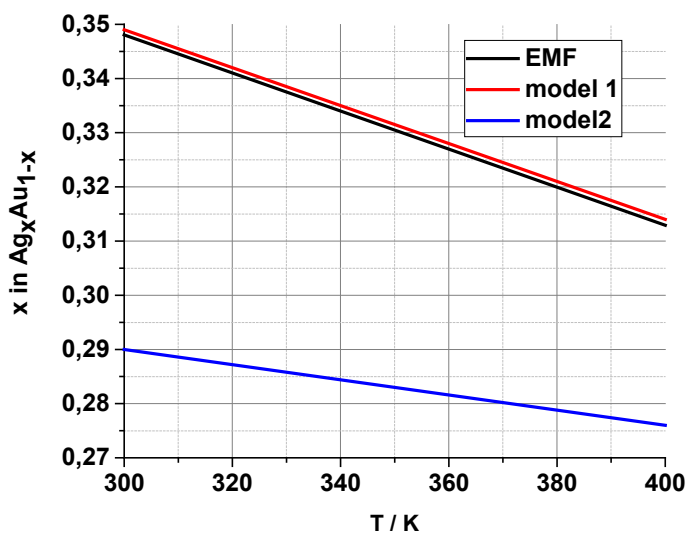

**Fig.2.** The values of the composition of the $Ag_xAu_{1-x}$ alloy in equilibrium with $Ag_2S$-$Ag_3AuS_2$-$Ag_xAu_{1-x}$ electromotive force (EMF). The EMF values are obtained from Table 1. Model 1 represents the values obtained by modeling in the TernApi program using R1. Model 2 represents the values obtained by modeling in the TernApi program using the reactions given in the article [1].

It should be noted that White's data [5] were chosen as an example, and it was also found that when correctly calculating thermodynamic systems, the choice of thermodynamic model for the Ag-Au alloy does not make a significant difference (the resultant figure in Fig.2 shifts up or down without undergoing significant changes), and the final result remains within the error range of the calculations for obvious reasons.

Literature.